\documentclass[referee]{raa}

\usepackage{graphicx,times}
\usepackage{multirow} 
\usepackage{natbib}
\usepackage{amssymb,amsmath}
\usepackage{xcolor}
\bibpunct{(}{)}{;}{a}{}{,}

\usepackage[pagebackref=true]{hyperref}

\begin{document}

   \title{Measurement of the Weyl Potential Evolution and $E_G$ Statistic from KiDS-1000, BOSS and 2dFLenS}

 \volnopage{ {\bf 20XX} Vol.\ {\bf X} No. {\bf XX}, 000--000}
   \setcounter{page}{1}

   \author{Xu-Wei Zhang 
   \inst{1,2}, Ming Zhang\inst{1,2*}, Yun-Liang Ren\inst{3}, Xiao-Feng Yang\inst{4}
   \footnotetext{$*$Corresponding Author}
   }

   \institute{State Key Laboratory of Radio Astronomy and Technology, Xinjiang Astronomical Observatory, Chinese Academy of Sciences, Urumqi 830011, China; {\it zhangxuwei23@mails.ucas.ac.cn}\\
\and
School of Astronomy and Space Science, University of Chinese Academy of Sciences, Beijing 100049, China\\
\and
School of Physical Science and Technology, Xinjiang University, Urumqi, Xinjiang 830046, China\\
\and
School of Physics and Electronics, Henan University, Kaifeng, Henan 475004, China\\
\vs \no
   {\small Received 20XX Month Day; accepted 20XX Month Day}
}

\abstract{A recently developed model-independent approach to measuring the Weyl potential has shown some tensions with $\Lambda$CDM (Lambda Cold Dark Matter) in DES (Dark Energy Survey) Y3 data. We apply this framework to Kilo-Degree Survey (KiDS-1000) weak lensing and BOSS/2dFLenS galaxy clustering using the KiDS Cosmology Analysis Pipeline (KCAP) in two redshift bins. To apply this method to the BOSS spectroscopic wedges, we take the early-time power spectrum from a redshift where General Relativity (GR) is recovered and jointly fit the late-time observables $\hat{b}$, $\hat{f}$, and $\hat{J}$. With Planck18 priors, the CMASS measurement of $\hat{J}$ is $1.50\sigma$ below the GR prediction, while the LOWZ measurement remains consistent with it. The decrease in the central value of $\hat{J}$ from LOWZ to CMASS provides a mild indication of the previously reported ``Lensing is low'' behavior. Without Planck18 priors, we obtain $E_G(0.38)=0.503\pm0.054$ and $E_G(0.61)=0.444\pm0.054$, both consistent with GR. Furthermore, phenomenological gravity tests also show no significant preference for modified gravity.
}

   \authorrunning{X.-W. Zhang et al. }            
   \titlerunning{Weyl Potential and $E_G$ from KiDS and BOSS}  
   \maketitle

%
\section{Introduction}
$\Lambda$CDM (Lambda Cold Dark Matter) has achieved remarkable success in explaining a wide range of observations \citep{Weinberg_2013,Bull_2016,Joyce_2015}. It attributes the accelerated expansion of the universe to the cosmological constant ($\Lambda$), and the formation of cosmic structures to cold collisionless dark matter \citep{Peebles_2003,Frieman_2008}. However, the fundamental nature of these two components remains unknown. Furthermore, the increasing precision of recent cosmological observations has revealed several significant tensions, most notably the $H_0$ tension between the expansion rate measurements in the early and late universe \citep{Riess_2019,Verde_2019,Di_Valentino_2021,Knox_2020}, the $S_8$ tension regarding the amplitude of matter clustering \citep{Heymans_2021,Abbott_2022}, and emerging hints of dynamical dark energy. These discrepancies suggest that either unaccounted systematic errors exist in those observations, or the $\Lambda$CDM paradigm itself may be in danger of breaking down \citep{Bull_2016}. To robustly test this model, it is therefore useful to scrutinize several assumptions that enter its construction, including the validity of General Relativity (GR) on cosmological scales, the Friedmann-Lemaître-Robertson-Walker (FLRW) description implied by large-scale homogeneity and isotropy, and the nature of the dark components.

The large-scale structure (LSS) of the universe provides a powerful arena for testing these paradigms \citep{Percival_2009,Clifton_2012,Koyama_2016}. Nevertheless, such studies face two major challenges. First, the theoretical landscape beyond $\Lambda$CDM is extraordinarily vast. It encompasses various modified gravity, dynamical dark energy, and interacting dark matter models, making it computationally and practically unfeasible to test each model individually. Second, strong observational degeneracies exist between different theoretical frameworks. For example, a fifth force acting on dark matter would modify the Poisson equation, and its effects on structure growth, when probed solely through galaxy surveys, might be observationally indistinguishable from modifications to gravity \citep{Baker_2015,Pogosian_2016}. Therefore, developing observational tests capable of breaking these degeneracies and isolating specific physical signatures of physics beyond $\Lambda$CDM has become crucial \citep{bonvin2025noveltestgravitydoes}.

To address these challenges, two complementary approaches are widely adopted. The first method utilizes generalized phenomenological frameworks, such as Horndeski theories or the Effective Field Theory of Dark Energy (EFTofDE) \citep{Gleyzes_2013,Bloomfield_2013,Bellini_2014}. These frameworks, built on the symmetries of the Lagrangian, encompass a broad class of modified gravity theories, allowing a large number of specific models to be simultaneously ruled out by their parameter space constraints \citep{Clifton_2012,Koyama_2016}. The second approach focuses on the observational side, aiming to identify and measure observables that can capture distinct gravitational signatures independent of specific models. Various such tests have been proposed and implemented, including tests of the cosmic distance duality relation \citep{wang_testing_2024,zhang_testing_2026}, tests of the Cosmological Principle and the associated FLRW description \citep{Aluri_2023,Secrest_2022,chen_testing_2026}, tests of the Weak Equivalence Principle (WEP) \citep{Bonvin_2018}, and tests of the scale invariance of structure growth \citep{franco_null_2020}. Additionally, consistency checks for the $\Lambda$CDM model have been developed, for instance, via evolution equations \citep{arjona2024probinglambdacdmweylpotential} or the $E_G$ statistic \citep{zhang_probing_2007,grimm_testing_2024}.

Recently, analogous to the theoretical developments that enable the measurement of the growth observable $\hat{f}(z) \equiv f(z)\sigma_8(z)$ from redshift space distortions (RSD), a similar formalism has been extended to weak gravitational lensing observables \citep{tutusaus_combining_2023}. This theoretical framework provides a direct measurement of the Weyl potential that is independent of theoretical models. Following this, preliminary measurements and subsequent tests of gravity theories were conducted \citep{tutusaus_measurement_2024}. They found that in the two lower redshift bins of the DES Y3 2x2pt analysis, the Weyl potential was lower than the $\Lambda$CDM predictions, and these lower values were interpreted as a contributor to the discrepancy in the matter clustering parameter $\sigma_8$ between CMB and weak lensing measurements. This is also related to recent discussions of whether $S_8$ and $f\sigma_8$ constraints show redshift-dependent trends, with possible relevance for ISW probes of the time evolution of large-scale potentials \citep{Adil_2023,Akarsu_2024_tensions,Akarsu_2025_growth}. In our work, we consider the joint analysis of the KiDS and BOSS samples. Our goal is to obtain similar Weyl potential measurements in this dataset, update the results of the $E_G$ statistic and gravity tests using this model-independent approach, and compare them with findings from previous studies.

\section{Methodology and Data}
\label{sect:the}
\subsection{Observables Encoded in the Weyl Potential}
Following \citet{tutusaus_combining_2023,tutusaus_measurement_2024}, we will first briefly review the Weyl-potential formalism below.
Consider the evolution of cosmological perturbations in a flat FLRW universe. In the conformal Newtonian gauge, the perturbed line element is expressed as:
\begin{equation}
    ds^2 = a^2(\tau) \left[ -(1+2\Psi) d\tau^2 + (1-2\Phi) d\mathbf{x}^2 \right] \,,
\end{equation}
where $\Psi$ and $\Phi$ are the time and spatial metric potentials, respectively. Gravitational lensing is sensitive to their combination, usually written as the Weyl potential (or lensing potential) $\Psi_W = (\Phi + \Psi)/2$. In GR and in the absence of anisotropic stress, the Einstein equations give $\Phi = \Psi$, so the Weyl potential reduces to the common metric potential. The Poisson equation then relates this potential to the matter density contrast $\delta$ via:
\begin{equation}
    -k^2 \Psi_W(k,z) = \frac{3}{2} \mathcal{H}^2(z) \Omega_m(z) \delta(k,z) \,,
\end{equation}
where $\mathcal{H}(z)$ is the conformal Hubble parameter and $\Omega_m(z)$ is the matter density parameter. Assuming linear structure growth, the density contrast evolves as $\delta(k,z) =[D_1(z)/D_1(z_*)] \delta(k,z_*)$, where $D_1(z)$ is the linear growth factor, and $z_*$ is an initial redshift within the matter-dominated era.

To accurately model observables, the non-linear matter power spectrum at redshift $z$, $P^{\text{nl}}_{\delta\delta}(k,z)$ must be related to the primordial linear spectrum. So we can decompose it into a linear growth term and a non-linear boost factor $B(k,z)$ as follows:
\begin{equation}
    P^{\text{nl}}_{\delta\delta}(k,z) = \left[ \frac{D_1(z)}{D_1(z_*)} \right]^2 P^\text{lin}_{\delta\delta}(k,z_*) B(k,z) \,,
\end{equation}
where $P^\text{lin}_{\delta\delta}(k,z_*)$ is the linear matter power spectrum evaluated at the initial redshift $z_*$ which is chosen at a high redshift where we assume GR is restored {(we will set $z_* = 5$ in this work)}. Substituting this decomposition into the Limber integrals, the standard GR angular power spectra for galaxy-galaxy lensing (GGL), $C_\ell^{\Delta \kappa}$ (where $\Delta$ denotes the projected galaxy overdensity, $\kappa$ is the lensing convergence), and galaxy clustering (GC), $C_\ell^{\Delta \Delta}$, are given by:
\begin{align}
    C_\ell^{\Delta \kappa}(z_i,z_j) &= \frac{3}{2} \int dz \, n_i(z) \mathcal{H}^2(z) \Omega_m(z) b_i(z) \left[ \frac{D_1(z)}{D_1(z_*)} \right]^2 \nonumber \\
    &\quad \times P^\text{lin}_{\delta\delta}(k_\ell, z_*) B(k_\ell, z) \int dz' \, n_j(z') \frac{\chi(z')-\chi(z)}{\chi(z)\chi(z')} \,, \label{eq:Cl_GG_GR} \\
    C_\ell^{\Delta \Delta}(z_i,z_j) &= \int dz \, n_i(z) n_j(z) \frac{\mathcal{H}(z)(1+z)}{\chi^2(z)} b_i(z)b_j(z) \left[ \frac{D_1(z)}{D_1(z_*)} \right]^2 \nonumber \\
    &\quad \times P^\text{lin}_{\delta\delta}(k_\ell, z_*) B(k_\ell, z) \,, \label{eq:Cl_GC_GR}
\end{align}
where $b_i(z)$ is the linear galaxy bias, $\chi$ is the comoving distance, and $k_\ell = (\ell+1/2)/\chi(z)$.

To test theories of gravity beyond GR, phenomenological functions $\mu$ and $\Sigma$ are conventionally introduced to parameterize modifications to the Poisson equations for $\Psi$ and $\Psi_W$, respectively:
\begin{align}
    -k^2 \Psi &= \frac{3}{2} \mathcal{H}^2 \Omega_m \mu(k,z) \delta \,, \label{eq:possion1}\\
    -k^2 \Psi_W &= \frac{3}{2} \mathcal{H}^2 \Omega_m \Sigma(k,z) \delta \,.\label{eq:possion2}
\end{align}
In GR, $\mu$ and $\Sigma$ are equal to 1. However, modified gravity theories often predict these functions to be redshift- and scale-dependent. To mitigate this model dependence, we can introduce a generalized Weyl evolution function $J(k,z)$, defined such that $-k^2 \Psi_W \propto J(k,z)\,\delta(k,z_*)/D_1(z_*)$. Although $J$ is in principle scale-dependent, the current data especially the weak lensing measurements used here do not support a robust scale-dependent analysis, because small-scale lensing and clustering are coupled to baryonic effects and scale-dependent galaxy bias. To further decouple the clustering amplitude from the primordial spectrum, two dimensionless observables, the Weyl evolution $\hat{J}$ and galaxy bias $\hat{b}_i$, are introduced:
\begin{align}
    \hat{J}(k, z) &\equiv \frac{J(k, z) \sigma_8(z)}{D_1(z)} = \frac{J(k, z) \sigma_8(z_*)}{D_1(z_*)} \,, \\
    \hat{b}_i(z) &\equiv b_i(z) \sigma_8(z) = b_i(z) \sigma_8(z_*) \frac{D_1(z)}{D_1(z_*)} \,.
\end{align}
Here, $\sigma_8(z_*)=\sigma_8 D_1(z_*)/D_1(0)$ is the clustering amplitude at redshift $z_*$. In GR, the evolution of the Weyl potential is governed by the background density and linear growth, yielding $J_{\text{GR}}(z) = \Omega_m(z) D_1(z)$. Therefore, $\hat{J}_{\text{GR}}(z) = \Omega_m(z) \sigma_8(z_*) D_1(z)/D_1(z_*)$. 

Based on these definitions, the angular power spectra can be written in a model-independent form that only requires the restoration of GR at high redshifts:
\begin{align}
    C_\ell^{\Delta \kappa}(z_i,z_j) &= \frac{3}{2} \int dz \, n_i(z) \mathcal{H}^2(z) \hat{b}_i(z) \hat{J}(k_\ell, z) B(k_\ell, z) \frac{P^\text{lin}_{\delta \delta}(k_\ell, z_*)}{\sigma_8^2(z_*)} \int dz' \, n_j(z') \frac{\chi(z')-\chi(z)}{\chi(z)\chi(z')} \,, \label{eq:Cl_GG_Final} \\
    C_\ell^{\Delta \Delta}(z_i,z_j) &= \int dz \, n_i(z) n_j(z) \frac{\mathcal{H}(z)(1+z)}{\chi^2(z)} \hat{b}_i(z) \hat{b}_j(z) B(k_\ell, z) \frac{P^\text{lin}_{\delta\delta}(k_\ell, z_*)}{\sigma_8^2(z_*)} \,. \label{eq:Cl_GC_Final}
\end{align}
Under this framework, $\hat{J}$ effectively captures the potential modifications to the Poisson equation and anisotropic stress, while the joint analysis of $C_\ell^{\Delta \kappa}$ and $C_\ell^{\Delta \Delta}$ helps break the degeneracy between galaxy bias and modified gravity signals. By contrast, extending this construction to the lensing auto-correlation, usually referred to as cosmic shear (CS), would require integrating the lensing kernel continuously from the observer to the sources, because its angular power spectrum, $C_{\ell}^{\kappa \kappa}$, is not naturally split into the same localized redshift bins. That would in turn require adopting a functional form or interpolation scheme for the evolution of $\hat{J}$, which would reintroduce the model dependence that we are trying to avoid.

However, the above derivation holds for photometric surveys like DES. For 3D spectroscopic surveys like BOSS, equation (\ref{eq:Cl_GC_Final}) is no longer applicable within the KiDS Cosmology Analysis Pipeline (KCAP). Due to different data processing methods, the observable is not the galaxy clustering angular power spectrum $C_\ell^{\Delta \Delta}$, but rather the clustering wedges $\xi(s, \mu, z)$ which comes from Galilean-invariant Renormalised Perturbation Theory (gRPT) \citep{PhysRevD.73.063519,Salazar_Albornoz_2017,2020PhRvD.102j3530E}. Consequently, the power spectrum is not simply linear in the matter power spectrum as $P_{\text{gg}}\sim b^2P_{\delta \delta}$, but instead takes a more complex form:
\begin{equation}
    P_{gg,s}(k, \mu, z) = W\bigl(k\mu f(z)\bigr) \left[ P_{gg} + 2 f(z) \mu^2 P_{g\theta} + f^2(z) \mu^4 P_{\theta \theta} + \text{higher-order terms} \right] \,,
    \label{eq:grpt_wedges}
\end{equation}
where $\mu$ is the cosine of the angle between the wavevector and the line of sight, and $f(z) = d\ln D/d\ln a$ is the linear structure growth rate. The damping function $W(x)$ is used to fit the Fingers-of-God effect caused by virialized random motions at small scales. The terms $P_{gg}$, $P_{g\theta}$, and $P_{\theta\theta}$ represent the galaxy autopower spectrum, the galaxy velocity divergence cross spectrum, and the velocity divergence autopower spectrum, respectively. In the gRPT description these quantities are highly nonlinear; the galaxy sector is additionally controlled by two local biases (linear bias $b_1$, quadratic bias $b_2$) and one nonlocal bias $\gamma_3$.

{For the spectroscopic wedges, the early-time power spectrum and late-time observables are separated before the likelihoods are evaluated. At $z_\star=5$ we construct the early-time spectrum
\begin{equation}
    Q_\star(k) \equiv \frac{P^{\rm lin}_{\delta\delta}(k,z_\star)}{\sigma_8^2(z_\star)} \,.
    \label{eq:qstar}
\end{equation}
For each spectroscopic bin we independently sample
\begin{equation}
    \hat{b}_i=b_i\sigma_8(z_i), \qquad
    \hat{f}_i=f_i\sigma_8(z_i), \qquad
    \hat{J}_i \,.
    \label{eq:late_observables}
\end{equation}
In the large-scale linear limit of equation~\eqref{eq:grpt_wedges}, $W\to1$, while $P_{gg}=b_i^2P_{\delta\delta}^{\rm lin}$, $P_{g\theta}=b_iP_{\delta\delta}^{\rm lin}$, and $P_{\theta\theta}=P_{\delta\delta}^{\rm lin}$. The bracket therefore becomes $(b_i+\mu^2f_i)^2P_{\delta\delta}^{\rm lin}$ \citep{Kaiser_1987,Percival_2009}. Since $P_{\delta\delta}^{\rm lin}/\sigma_8^2$ is invariant under linear growth and equals $Q_\star$ in equation~\eqref{eq:qstar}, the definitions in equation~\eqref{eq:late_observables} give
\begin{align}
    P^{s,{\rm lin}}_{gg}(k,\mu,z_i) &\propto
    \left(\hat{b}_i+\mu^2\hat{f}_i\right)^2 Q_\star(k) \, ,
    \label{eq:zstar_boss}\\
    C_\ell^{\Delta\kappa}(z_i,z_j) &\propto
    \hat{b}_i\hat{J}_i Q_\star(k_\ell) \, .
    \label{eq:zstar_ggl}
\end{align}
The second relation follows from equation~\eqref{eq:Cl_GG_Final} after inserting equation~\eqref{eq:qstar} and taking the same large-scale limit, $B\to1$; the unchanged geometrical kernels are suppressed. Thus, the overall and anisotropic parts of the BOSS wedges constrain $\hat{b}_i$ and $\hat{f}_i$, respectively. Galaxy--galaxy lensing first constrains the product $\hat{b}_i\hat{J}_i$; because the same $\hat{b}_i$ is used in both likelihoods, their combination separates $\hat{J}_i$.

The gRPT calculation is evaluated with $b_i^{\rm eff}=\hat{b}_i/\sigma_8(z_i)$ and $f_i^{\rm eff}=\hat{f}_i/\sigma_8(z_i)$, together with $Q_\star$. The $\sigma_8$ then cancels from the leading linear terms. For the full galaxy--galaxy lensing calculation, $B$ is retained. Since the KCAP projection retains the $\Omega_m(z)$ prefactor in equation~\eqref{eq:Cl_GG_GR}, we match $\Omega_mP_{gm}^{\rm eff}=BQ_\star$, with $B=P_{\delta\delta}^{\rm nl}/P_{\delta\delta}^{\rm lin}$, giving
\begin{equation}
    P_{gm}^{\rm eff}(k,z)=
    \frac{P_{\delta\delta}^{\rm nl}(k,z)}{P_{\delta\delta}^{\rm lin}(k,z)}
    \frac{Q_\star(k)}{\Omega_m(z)} \, ,
    \label{eq:zstar_pgm}
\end{equation}
and the projected lensing term is multiplied by $\hat{b}_i\hat{J}_i$. The intrinsic-alignment and magnification contributions follow the KCAP model.

This construction extends the Tutusaus et al. method to spectroscopic wedges. The gRPT loop and nonlinear-bias terms, the nonlinear boost $P^{\rm nl}_{\delta \delta}/P^{\rm lin}_{\delta \delta}$, intrinsic alignment, magnification and nonlinear galaxy-bias correction continue to depend on the CAMB spectra and GR-based nonlinear modeling. The leading linear dependence on $\sigma_8$ is removed, although some nonlinear dependence remains.}
\subsection{\texorpdfstring{The $E_G$ Statistic}{The EG Statistic}}
Additionally, the $E_G$ statistic is also an essential observable for probing gravitational slip and testing gravity theories, and in a projected analysis it can be written as \citep{zhang_probing_2007}:
\begin{equation}
    E_G(\ell,z)=\Gamma(z) \frac{C_\ell^{\Delta \kappa}(z)}{\beta(z) C_\ell^{\Delta \Delta}(z)}
\end{equation}
where $\beta(z)=f(z)/b(z)$ is the redshift space distortion parameter and $\Gamma(z)$ is the geometry-dependent prefactor of the projected estimator. In our analysis, because we directly measure $\hat{f}(z) \equiv f(z)\sigma_8(z)$ and $\hat{J}(z)$, $E_G$ can be evaluated equivalently through their ratio \citep{grimm_testing_2024}:
\begin{equation}
    E_G=\left(\frac{\mathcal{H}(z)}{\mathcal{H}_0}\right)^2\frac{1}{1+z}\frac{\hat{J}(z)}{\hat{f}(z)}
\end{equation}
where $\mathcal{H}_0 \equiv \mathcal{H}(z=0)$. {We evaluate this expression for every posterior sample. Equivalently, the prefactor is $E^2(z)/(1+z)^3$, where $E(z)=H(z)/H_0$. The factor $\sigma_8(z)$ cancels in $\hat{J}/\hat{f}$. So $E_G$ remains suitable for a No-Prior GR consistency test even though the two observables cannot be compared separately with their absolute GR predictions.} We also include historical measurements of $E_G$ from the literature in Table~\ref{tab:EG_compilation}.
\begin{table}[htbp]
    \centering
    \caption{Historical measurements of the $E_G$ statistic from the literature.}
    \label{tab:EG_compilation}
    \begin{tabular}{ccc}
        \hline
        \textbf{Redshift} & \textbf{$E_G$} & \textbf{Reference} \\
        \hline
        0.37 & $0.392 \pm 0.065$ & \citealt{Reyes_2010} \\
        0.32 & $0.48 \pm 0.10$  & \citealt{blake_2-degree_2016} \\
        0.57 & $0.30 \pm 0.07$  & \citealt{blake_2-degree_2016} \\
        0.27 & $0.43 \pm 0.13$  & \citealt{Amon_2018} \\
        0.31 & $0.27 \pm 0.08$  & \citealt{Amon_2018} \\
        0.55 & $0.26 \pm 0.07$  & \citealt{Amon_2018} \\
        0.57 & $0.420 \pm 0.056$ & \citealt{Alam_2016} \\
        0.57 & $0.24 \pm 0.06$  & \citealt{Pullen_2016} \\
        0.316 & $0.40^{+0.11}_{-0.09}$ & \citealt{PhysRevD.109.083540} \\
        0.555 & $0.36^{+0.06}_{-0.05}$ & \citealt{PhysRevD.109.083540} \\
        0.295 & $0.447\pm 0.027$ & \citealt{grimm_testing_2024} \\
        0.467 & $0.378\pm 0.026$ & \citealt{grimm_testing_2024} \\
        0.626 & $0.396\pm 0.033$ & \citealt{grimm_testing_2024} \\
        0.771 & $0.345\pm 0.039$ & \citealt{grimm_testing_2024} \\
        \hline
    \end{tabular}
\end{table}

\subsection{Tests of Modified Gravity Theories}
To investigate the Weyl potential evolution $\hat{J}$ under modified gravity frameworks, we use the $\mu$-$\Sigma$ parameterization for modeling. From the Poisson equations \eqref{eq:possion1} and \eqref{eq:possion2} discussed previously, we obtain the parameterization as:
\begin{equation}
    \hat{J}(z)=\Sigma(z)\Omega_m(z)D_1(z)\frac{\sigma_8(z_*)}{D_1(z_*)}
\end{equation}
The advantage of this approach is that we do not need to implement specific gravity models directly into the pipeline. Once we obtain a set of $\hat{J}$ values, we can easily use them to test theories without running a full sampling pipeline for every individual model.

For a phenomenological test, we fix $\mu = 1$ and infer $\Sigma$ through $\hat{J}$. We consider three different choices for time evolution following \citet{tutusaus_measurement_2024}, formalized as $\Sigma(z)=1+\Sigma_0g(z)$, using the following models: 1) Standard evolution model: $g(z) = \Omega_{\Lambda}(z)$; 2) Constant (no evolution) model: $g(z) = 1$ for $z \in [0, 1]$ and $0$ otherwise; 3) Exponential evolution model: $g(z) = \exp(1 + z)$ for $z \in [0, 1]$ and $0$ otherwise. The first model is motivated by the idea that deviations from GR are associated with cosmic acceleration and are expected to decay as the dark energy fraction decreases (which is widely used in DES and Planck analyses \citep{Abbott_2022,planck_2020}). The second and third models are not physically motivated but only to explore the sensitivity of the data.

Since these models are linear, we use the least squares method for parameter estimation and error derivation. Fixing the background cosmological parameters $(\Omega_{m,0}, \sigma_{8,0})$, the Weyl potential evolution model $\hat{J}(z) = [1+\Sigma_0 g(z)]\hat{J}_{\Lambda\mathrm{CDM}}(z)$ depends linearly on the parameter $\Sigma_0$. We define the observational residual vector $\boldsymbol{\Delta} = \hat{\mathbf{J}}_{\mathrm{obs}} - \hat{\mathbf{J}}_0$ and the design matrix $\mathbf{X} = g(\mathbf{z}_i)\odot\hat{\mathbf{J}}_0$, where $\hat{\mathbf{J}}_0 \equiv \hat{J}_{\Lambda\mathrm{CDM}}(\mathbf{z}_i)$. Based on the inverse of the observational covariance matrix $\mathbf{C}^{-1}$, we construct the $\chi^2$ statistic:
\begin{equation}
\chi^2(\Sigma_0) = (\boldsymbol{\Delta}-\Sigma_0\mathbf{X})^{\mathsf{T}}\mathbf{C}^{-1}(\boldsymbol{\Delta}-\Sigma_0\mathbf{X}).
\end{equation}
Minimizing $\chi^2$ yields the analytical best fit solution for $\Sigma_0$ as $\hat{\Sigma}_0 = (\mathbf{X}^{\mathsf{T}}\mathbf{C}^{-1}\boldsymbol{\Delta})/(\mathbf{X}^{\mathsf{T}}\mathbf{C}^{-1}\mathbf{X})$, and its statistical error is given by the Fisher information matrix: $\sigma_{\Sigma_0} = (\mathbf{X}^{\mathsf{T}}\mathbf{C}^{-1}\mathbf{X})^{-1/2}$. This analytical method avoids numerical optimization, providing the minimum $\chi^2$ value and the significance of the deviation from standard gravity ($\Sigma_0=0$).

\subsection{Data}
\label{subsect:data}
The joint analysis in this study relies on weak lensing shear catalogs and spectroscopic galaxy redshift surveys. For background source galaxies, we use the weak lensing shear catalogs from the Kilo-Degree Survey's fourth data release, KiDS-1000 \citep{kuijken_fourth_2019,hildebrandt_kids-1000_2021}. The source galaxies in this dataset have redshifts estimated via multi-band photometry and are divided into five tomographic bins.

For foreground lens galaxies, we employ the Luminous Red Galaxy (LRG) sample from the Baryon Oscillation Spectroscopic Survey (BOSS) Data Release 12 \citep{alam_clustering_2017}. Following standard large-scale structure conventions, this sample is split into two independent redshift bins: the low redshift bin (LOWZ, the lower redshift BOSS LRG sample, $0.20 < z < 0.50$, effective redshift $z_{\rm eff} = 0.38$) and the high redshift bin (CMASS, the higher redshift BOSS constant-mass LRG sample, $0.50 < z < 0.75$, effective redshift $z_{\rm eff} = 0.61$).

In galaxy-galaxy lensing (GGL) measurements, to improve the signal-to-noise ratio and increase the overlap with the KiDS-South region, our analysis combines the BOSS samples with the LRG samples from the 2-degree Field Lensing Survey, 2dFLenS \citep{blake_2-degree_2016}. Because 2dFLenS uses identical color magnitude selection criteria to BOSS, it forms consistent foreground lens samples. However, for 3D galaxy clustering measurements, the small volume of the 2dFLenS sample limits its statistical contribution. Therefore, following the baseline data processing from previous studies \citep{blake_testing_2020}, we use the BOSS DR12 sample for clustering statistics (wedges measurements) and assume clustering characteristics are common with the full combined LRG samples.

To make the role of the CMB information explicit, we analyze two inference configurations. The ``No Prior'' configuration uses uniform priors on the cosmological parameters and no information from Planck. The ``With Prior'' configuration supplements it with Gaussian Planck18 priors, namely $\Omega_c h^2 = 0.1200 \pm 0.0012$, $\Omega_b h^2 = 0.02237 \pm 0.00015$, $h_0 = 0.6736 \pm 0.0054$, $n_s = 0.9649 \pm 0.0042$, and $S_8 = 0.832 \pm 0.013$. {Both configurations use the same KiDS source-redshift calibration priors, nuisance model, data vector, covariance, and scale cuts. We use the conservative GGL bandpower ranges $100\leq\ell\leq300$ for LOWZ and $100\leq\ell\leq600$ for CMASS, and the BOSS wedges over $20\leq s/(h^{-1}{\rm Mpc})\leq160$.}

Our results are obtained with the \texttt{CosmoSIS} software package\footnote{\url{https://github.com/cosmosis-developers/cosmosis}} \citep{zuntz_cosmosis_2015}, using CAMB\footnote{\url{https://github.com/cmbant/CAMB}} for the theoretical predictions and the Nautilus sampler \citep{lange_nautilus_2023}, a neural network based nested sampling module. We adopt $n_{\rm live}=500$ and $f_{\rm live}=0.01$, and perform the analysis in parallel with OpenMPI\footnote{\url{https://www.open-mpi.org/}}.

\section{Results}
\subsection{Main Measurement Results}
\begin{table}[htbp]
    \centering
    \caption{Constraints on cosmological parameters and derived statistics. We report the mean and 68\% confidence intervals for both ``No Prior'' and ``With Prior'' configurations. The top section presents the base cosmological parameters; the middle sections list {the galaxy bias, growth observable, and Weyl potential evolution}; the bottom sections show {the derived $E_G$ statistics and goodness-of-fit}. Bayesian evidence is also provided for reference.}
    \label{tab:parameter_constraints}
    \begin{tabular}{lccc}
        \hline
        \textbf{Parameter} & \textbf{Symbol} & \textbf{With Prior} & \textbf{No Prior} \\
        \hline
        \multicolumn{4}{c}{\textit{Base Cosmological Parameters}} \\
        \hline
        Matter density parameter & $\Omega_m$ & {$0.311\pm0.005$} & {$0.305\pm0.012$} \\
        Weak lensing parameter & $S_8$ & {$0.830\pm0.013$} & {$0.359\pm0.062$} \\
        Primordial amplitude & $A_s$ & {$(2.127\pm0.082)\times10^{-9}$} & {$(0.418\pm0.142)\times10^{-9}$} \\
        Spectral index & $n_s$ & {$0.964\pm0.004$} & {$0.897\pm0.040$} \\
        Hubble parameter & $h_0$ & {$0.677\pm0.005$} & {$0.692\pm0.021$} \\
        \hline
        \multicolumn{4}{c}{\textit{Galaxy Bias and Growth Observables}} \\
        \hline
        Galaxy bias (LOWZ) & $\hat{b}(z_1)$ & {$1.270\pm0.023$} & {$1.194\pm0.027$} \\
        Galaxy bias (CMASS) & $\hat{b}(z_2)$ & {$1.250\pm0.020$} & {$1.180\pm0.027$} \\
        Growth observable (LOWZ) & $\hat{f}(z_1)$ & {$0.512\pm0.039$} & {$0.487\pm0.037$} \\
        Growth observable (CMASS) & $\hat{f}(z_2)$ & {$0.442\pm0.032$} & {$0.415\pm0.031$} \\
        \hline
        \multicolumn{4}{c}{\textit{Weyl Potential}} \\
        \hline
        Weyl potential evolution (LOWZ) & $\hat{J}_1$ & {$0.371\pm0.041$} & {$0.428\pm0.035$} \\
        Weyl potential evolution (CMASS) & $\hat{J}_2$ & {$0.322\pm0.043$} & {$0.389\pm0.041$} \\
        \hline
        \multicolumn{4}{c}{\textit{Derived Statistics}} \\
        \hline
        $E_G$ statistic (LOWZ) & $E_G(z_1)$ & {$0.416\pm0.046$} & {$0.503\pm0.054$} \\
        $E_G$ statistic (CMASS) & $E_G(z_2)$ & {$0.348\pm0.048$} & {$0.444\pm0.054$} \\
        \hline
        \multicolumn{4}{c}{\textit{Goodness of Fit}} \\
        \hline
        Best-fit total $\chi^2$ ($N_{\rm data}=190$) & $\chi^2_{\rm min}$ & {$203.74$} & {$199.98$} \\
        Bayesian evidence & $\log \mathcal{Z}$ & {$-139.501\pm0.011$} & {$-135.665\pm0.011$} \\
        \hline
    \end{tabular}
\end{table}

\begin{figure}
    \centering
    \includegraphics[width=0.95\textwidth]{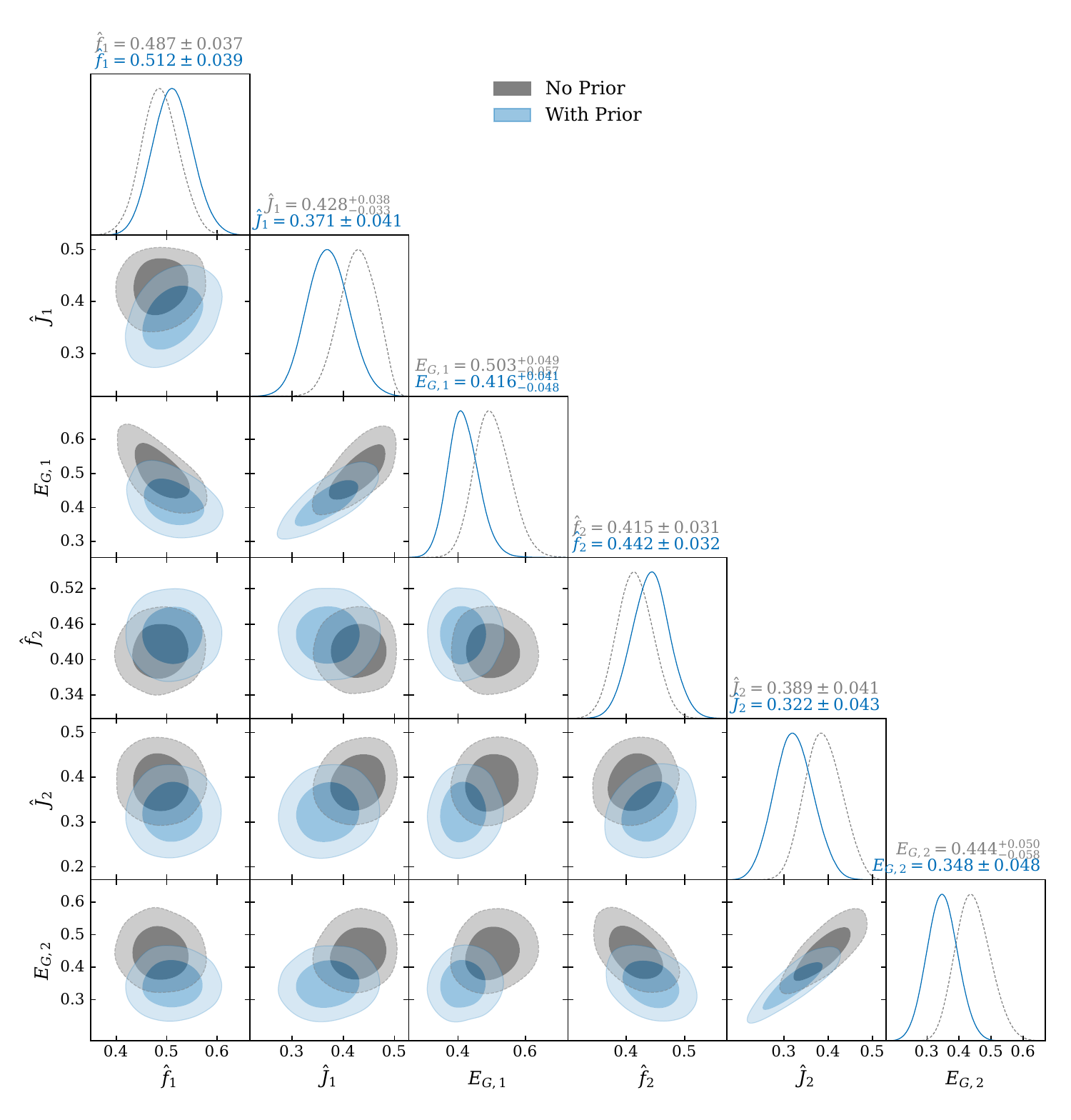}
    \caption{Marginalized posterior distributions for {$\hat{f}_i$, $\hat{J}_i$, and $E_{G,i}$} in the two redshift bins. Blue contours represent results with Planck priors, while grey contours represent results without Planck priors.}
    \label{fig:posterior_distribution}
\end{figure}

{Table~\ref{tab:parameter_constraints} and Figure~\ref{fig:posterior_distribution} summarize our constraints for the With-Prior and No-Prior configurations. The parameters $\hat{b}$ and $\hat{f}$ are varied directly in each BOSS bin, so their uncertainties include the freedom of the clustering response away from the GR growth history. The best-fit values are $\chi^2=203.74$ and $199.98$ for 190 data points in the With-Prior and No-Prior cases, respectively. The largest difference between the two configurations occurs in $\hat{b}_i$ (about $2\sigma$), the changes in $\hat{f}_i$ are below $0.7\sigma$, while those in $\hat{J}_i$ and $E_{G,i}$ are approximately $1.1$--$1.3\sigma$.}

\begin{figure}[htbp]
    \centering
    \includegraphics[width=0.49\textwidth]{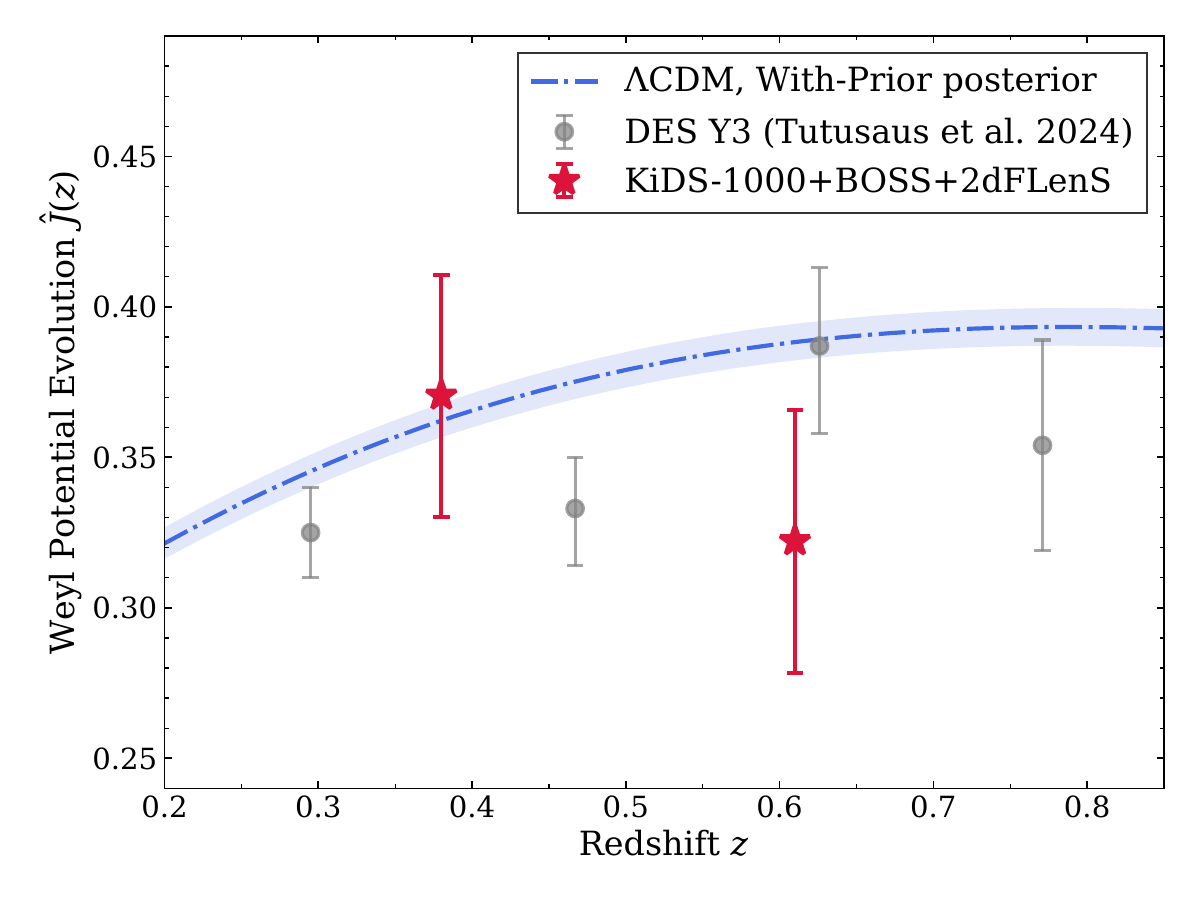}
    \includegraphics[width=0.49\textwidth]{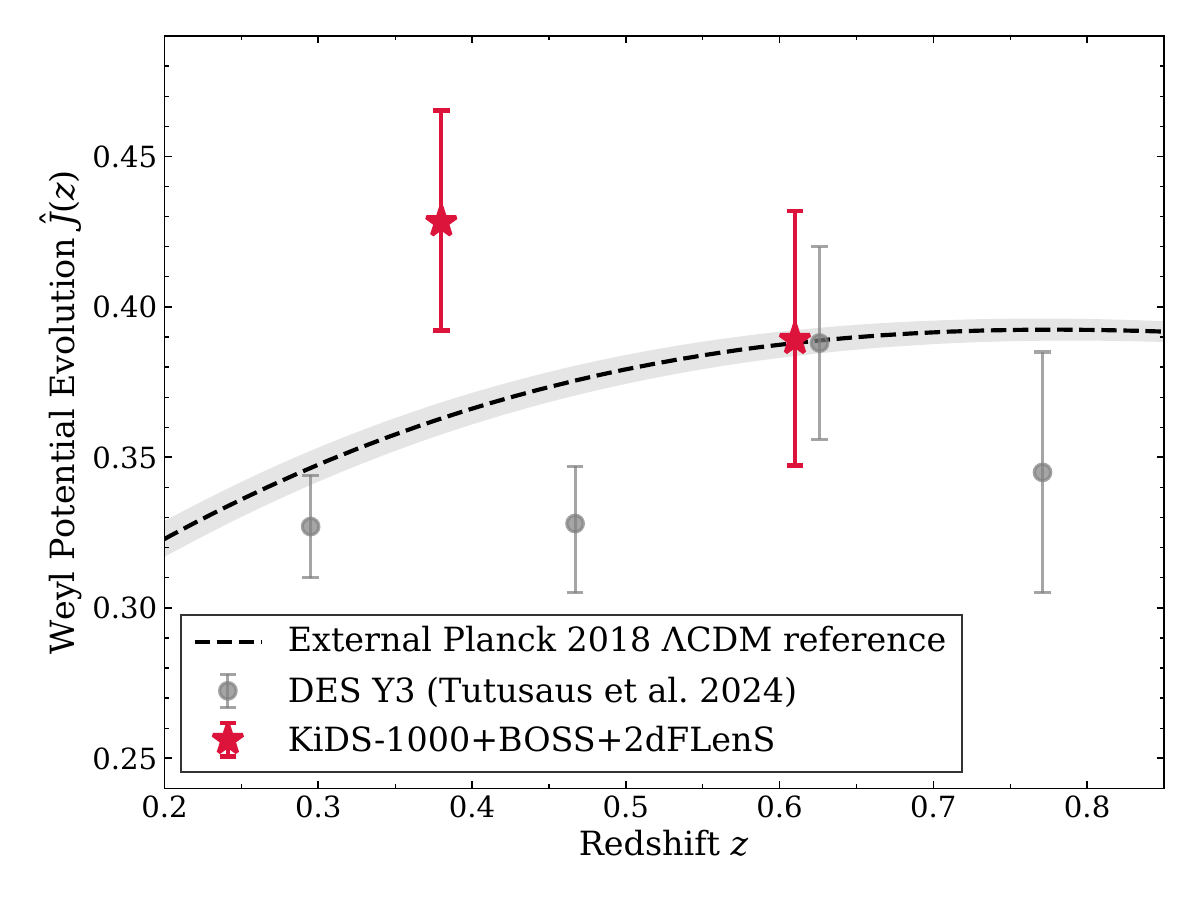}
    \caption{Measured values of $\hat{J}$ at the effective redshifts with 68\% credible intervals, comparing the With-Prior (left) and No-Prior (right) configurations. The red star points indicate our measurements, while the grey circles represent the DES Y3 measurements of \cite{tutusaus_measurement_2024}. {The left-panel curve is the GR prediction from the With-Prior posterior. The right-panel curve shows the Planck18 GR prediction only for comparison, since the No-Prior analysis cannot determine its own GR prediction.}}
    \label{fig:J_hat_evolution_comparison}
\end{figure}

{For the With-Prior chain, the GR predictions at $z=0.38$ and $0.61$ are $\hat{J}_{\rm GR}=0.362\pm0.006$ and $0.389\pm0.006$. The measured central values decrease from $0.371\pm0.041$ in LOWZ to $0.322\pm0.043$ in CMASS, whereas the GR prediction increases over the same redshift interval. The measurements differ from GR by $+0.20\sigma$ and $-1.50\sigma$, respectively. Thus, the suppression is mainly associated with the CMASS bin.

The No-Prior measurements also decrease from $0.428\pm0.035$ to $0.389\pm0.041$. Their absolute values cannot be used for an internal GR test because the primordial amplitude is not constrained in this configuration, and the Planck curve in the right panel is consequently shown only as an external reference. Nevertheless, the downward change in the two measured central values and the mildly low With-Prior CMASS point seems consistant with the ``Lensing is low'' behavior previously reported for this sample \citep{Pullen_2016,leauthaud_lensing_2017}. However, owing to only two redshift bins and a maximum deviation of $1.50\sigma$, this should be regarded as a mild indication rather than a statistically true redshift trend.}

{It should be noted that at linear order the BOSS wedges constrain $\hat{b}=b\sigma_8$ and $\hat{f}=f\sigma_8$ while GGL constrains $\hat{b}\hat{J}$. Fitting these combinations directly allows a change in the primordial amplitude to be compensated by changes in $b$, $f$ and $J$ without altering the leading order predictions, which means that the No-Prior $S_8$ and $A_s$ values are not physical amplitude constraints and mainly reflect residual nonlinear-model dependence together with nuisance degeneracies.}

\subsection{\texorpdfstring{Measurement Results for the $E_G$ Statistic}{Measurement Results for the EG Statistic}}
\begin{figure}[htbp]
    \centering
    \includegraphics[width=0.49\textwidth]{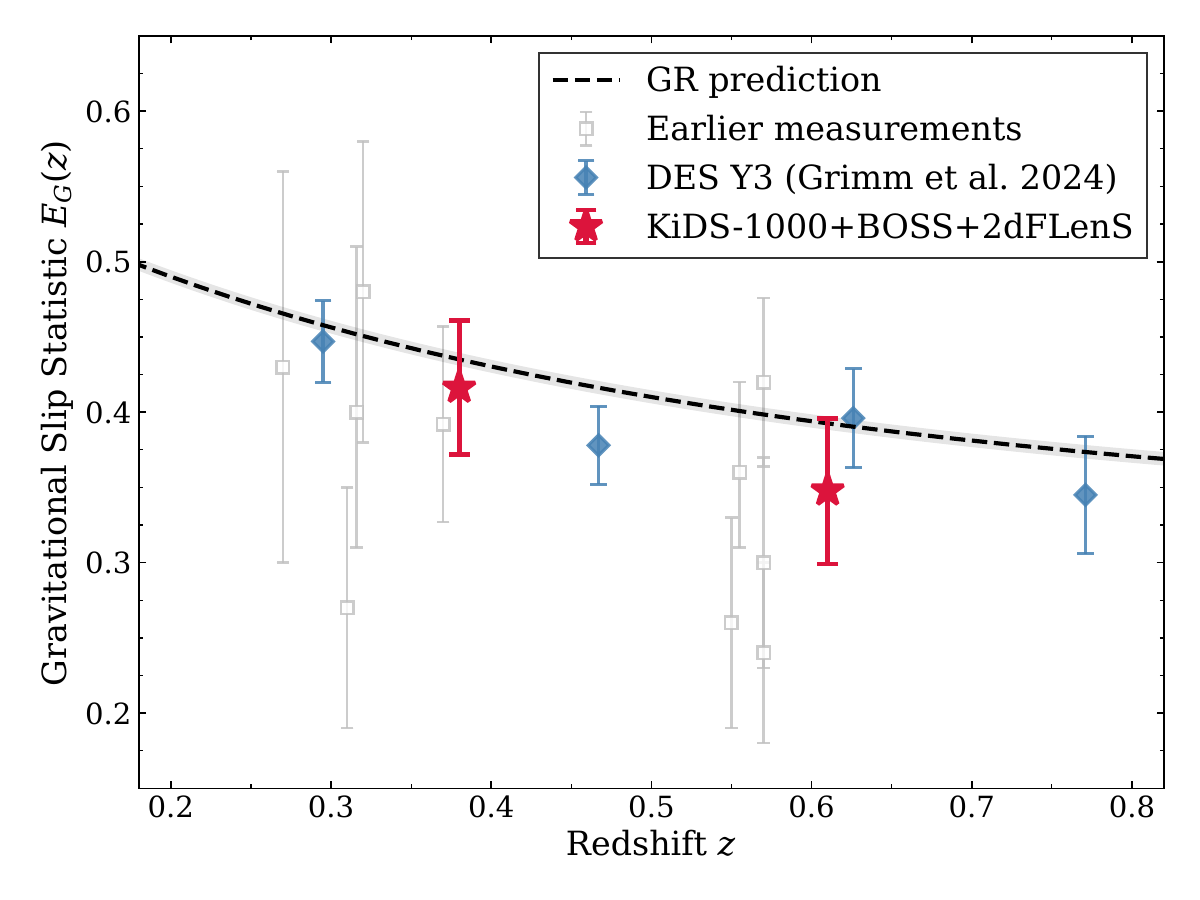}
    \includegraphics[width=0.49\textwidth]{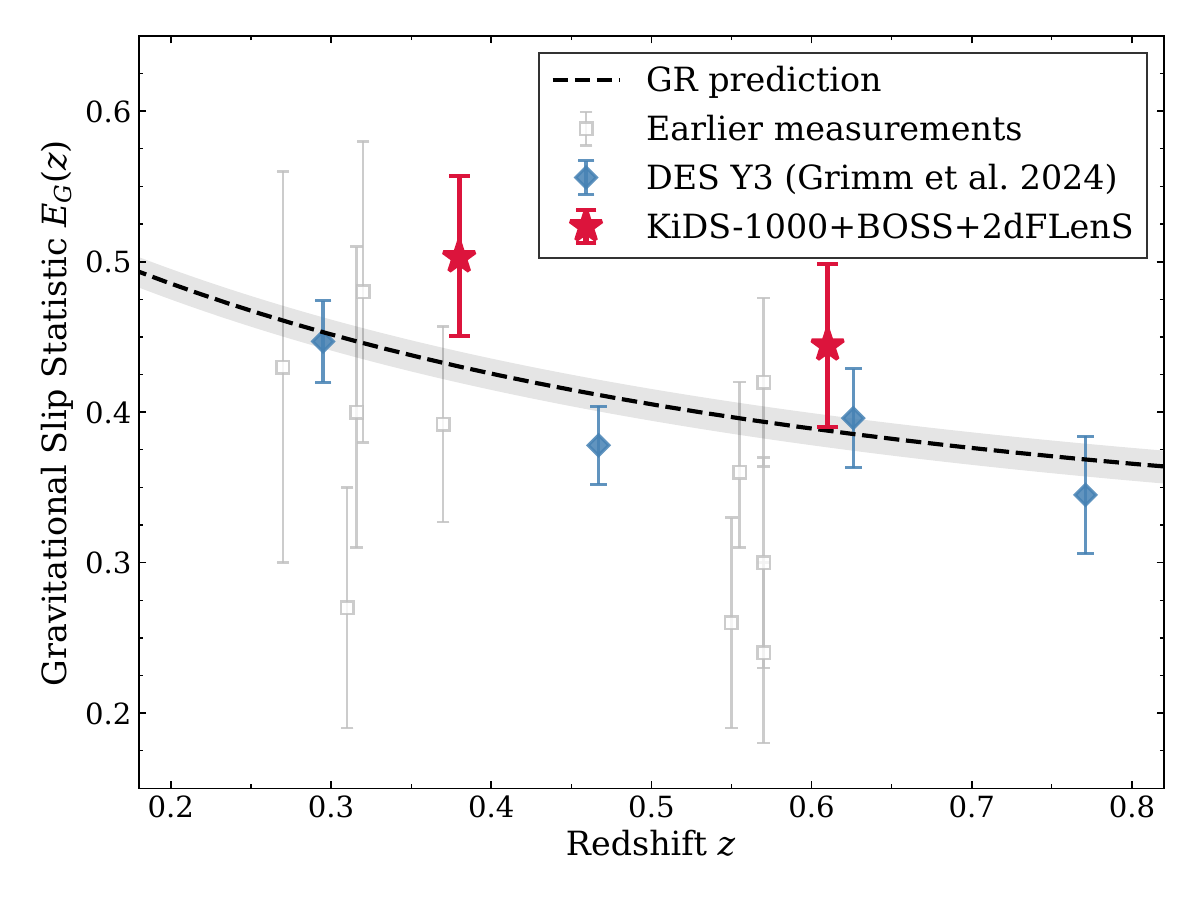}
    \caption{Measurements of $E_G$ (red star points) compared with published measurements for the With-Prior (left) and No-Prior (right) configurations.The dashed curves show the corresponding GR predictions and the grey shade is $1\sigma$ interval. The blue one show \citet{grimm_testing_2024}, and the remaining compilation is listed in Table~\ref{tab:EG_compilation}.}
    \label{fig:EG_compilation}
\end{figure}

{As shown in Figure~\ref{fig:EG_compilation}, the With-Prior results are $E_G(0.38)=0.416\pm0.046$ and $E_G(0.61)=0.348\pm0.048$. They differ from the conditional GR predictions by $-0.42\sigma$ and $-0.93\sigma$. The No-Prior values are $0.503\pm0.054$ and $0.444\pm0.054$, corresponding to $+1.34\sigma$ and $+1.05\sigma$ relative to the GR predictions based on the measured background expansion. Because $\sigma_8(z)$ cancels in $\hat{J}/\hat{f}$, the latter comparison does not require Planck information on the primordial power spectrum. All four measurements are therefore consistent with GR, and the data also provide no significant evidence for gravitational slip.}

\subsection{Constraints on Phenomenological Gravity Models}
\begin{figure}[htbp]
    \centering
    \includegraphics[width=0.49\textwidth]{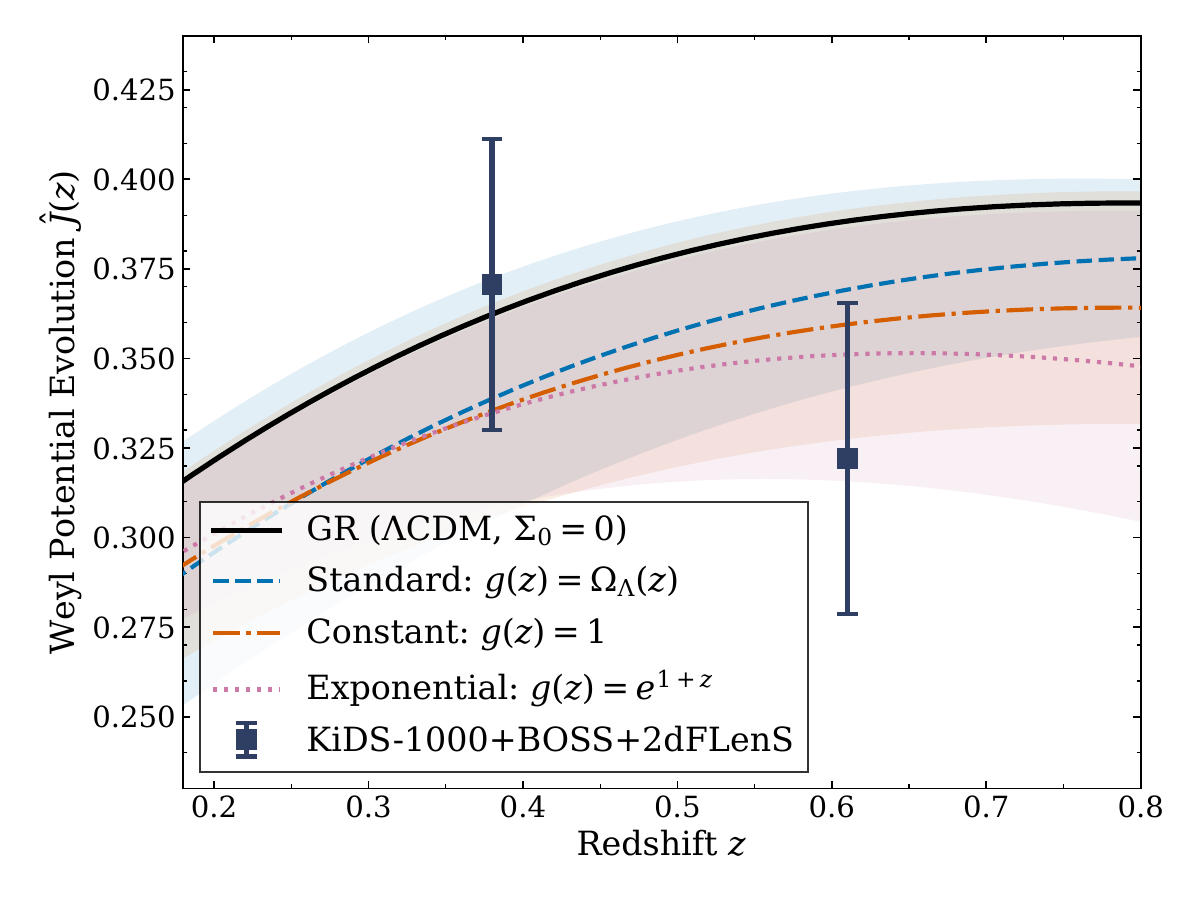}
    \caption{Predictions for $\hat{J}$ under modified gravity models with $\Sigma \neq 1$ and $\mu = 1$. {The background parameters $\Omega_{m,0}$ and $\sigma_{8,0}$ are fixed as With-Prior posterior means.} For the three evolution models (Standard, Constant, and Exponential), the colored shades indicate the $1\sigma$ uncertainty of the fit. The black line represents the baseline $\Lambda$CDM prediction.}
    \label{fig:J_hat_evolution_analytical}
\end{figure}

\begin{table}[htbp]
    \centering
    \caption{Parameter constraints and $\chi^2$ values for the baseline $\Lambda$CDM model and three phenomenological modified gravity models {for the With-Prior cosmology.}}
    \label{tab:mg_constraints}
    \begin{tabular}{lccc}
        \hline
        \textbf{Model} & \textbf{$\Sigma_0$} & \textbf{Deviation} & \textbf{$\chi^2$} \\
        \hline
        $\Lambda$CDM & -- & -- & {$2.44$} \\
        Standard & {$-0.142\pm0.203$} & {$0.70\sigma$} & {$1.95$} \\
        Constant & {$-0.074\pm0.083$} & {$0.90\sigma$} & {$1.63$} \\
        Exponential & {$-0.019\pm0.018$} & {$1.05\sigma$} & {$1.34$} \\
        \hline
    \end{tabular}
\end{table}

As illustrated in Figure~\ref{fig:J_hat_evolution_analytical} and Table~\ref{tab:mg_constraints}, in the With Prior case, {the fit of the baseline $\Lambda$CDM model gives $\chi^2=2.44$}. When $\Sigma_0$ is introduced, all three models yield negative values to suppress the predicted Weyl potential, with the exponential model yielding the lowest $\chi^2$. However, the deviations of $\Sigma_0$ from GR are only {about $0.70\sigma$ to $1.05\sigma$}. Given that the fit relies on only two data points and leaves 1 d.o.f., this level of deviation is also statistically insufficient to claim a breakdown of GR.

This trend contrasts with a previous study combining DES and BOSS CMASS samples \cite{lee_probing_2021}. Using the parameterization $\Sigma(z) = 1 + \Sigma_0 \frac{\Omega_{de}(z)}{\Omega_{de,0}}$, they reported a negative deviation from GR ($\Sigma_0 = -0.17^{+0.16}_{-0.15}$) using only late time data. Upon introducing Planck18 data, their $\Sigma_0$ shifted back to the GR limit ($\Sigma_0 = 0.078^{+0.078}_{-0.082}$). {Our Standard-model result, $\Sigma_0=-0.142\pm0.203$, instead retains a negative central value after applying the Planck prior, but its larger uncertainty makes it consistent with GR. This difference may reflect the use of different lens samples and observables and is not statistically significant.}

Compared to the reference analysis by \cite{tutusaus_measurement_2024}, our negative $\Sigma_0$ values align in magnitude and sign with their DES Y3 measurements ($-0.24 \pm 0.10$, $-0.13 \pm 0.06$, and $-0.027 \pm 0.013$). However, in their study, the standard $\Lambda$CDM model yielded poor fits overall. {In our With-Prior analysis, $\Lambda$CDM remains an adequate description of the two $\hat{J}$ measurements, and the reductions in $\chi^2$ after adding $\Sigma_0$ are not statistically significant.}

{But these are conditional $\mu=1$ constraints based on the two $\hat{J}$ measurements, rather than a joint $\mu$--$\Sigma$ inference. In the No-Prior configuration, $\hat{J}$ is degenerate with the primordial amplitude, so $\Sigma_0$ is reported only for the With-Prior configuration. It should be noted that the calculation retains some dependence on the GR model through the gRPT and nonlinear-power terms, as well as the intrinsic-alignment and magnification models.}

\section{Discussion and Conclusion}\label{sec:weyl_discuss_summary}
In summary, we use a model-independent method to measure the Weyl potential evolution $\hat{J}(z)$ and the $E_G(z)$ statistic, which probes gravitational slip, by combining KCAP with {a treatment of the BOSS spectroscopic wedges in which $\hat{b}$ and $\hat{f}$ are fitted independently}. Testing this on the KiDS-1000+BOSS+2dFLenS dataset, {we find that all GR consistency tests give deviations below $2\sigma$. The measured $\hat{J}$ decreases from LOWZ to CMASS in both configurations. When with Planck priors, the CMASS point is $1.50\sigma$ below GR, while LOWZ is consistent with GR. The decreasing trending shows a mild indication of the ``Lensing is low'' behavior associated with the CMASS sample research, although two redshift bins are insufficient to establish a significant evolution trend. Without Planck priors, $E_G(0.38)=0.503\pm0.054$ and $E_G(0.61)=0.444\pm0.054$ differ from GR by only $1.34\sigma$ and $1.05\sigma$. The phenomenological fits of $\Sigma_0$ also show no significant deviation from GR.}

Recent studies indicate that this ``Lensing is low'' problem may be related to systematics specific to this sample and modeling limitations. Using N-body simulations, \citet{luo_photometric_2025} showed that the stellar-mass completeness of the CMASS sample changes with redshift. This indicates that the samples require much finer tomographic binning instead of being treated as one broad bin. Similarly, \citet{chen_not_2024} found no such lensing suppression in the DESI LRG sample, indicating that the issue is specific to BOSS CMASS, and they also claim that ``Not all lensing is low''. Furthermore, \citet{Mahony_2025} applied a fully non-linear SHAMe model to the GAMA sample and successfully recovered an $S_8$ value consistent with Planck18. Together, these findings indicate that the deviation can be alleviated through better sample selection and more precise non-linear bias modeling. Future work should apply these advanced models to new datasets to further clarify this issue.

Moreover, future investigations of the Weyl potential and the $S_8$ tension will benefit from broader observational data and refined modeling. Cross survey analyses \citep{KiloDegreeSurvey2023gfr} and next generation weak lensing measurements from Euclid \citep{Euclid2022}, the Legacy Survey of Space and Time (LSST) \citep{Ivezi_2019}, and the Chinese Space Station Telescope (CSST) \citep{Gong_2019} will help isolate survey specific systematics. Additionally, CMB lensing constraints from the Simons Observatory \citep{Ade_2019} and updated hydrodynamical simulations \citep{Schaye_2023, mccarthy2023flamingoprojectrevisitings8} are expected to reduce theoretical uncertainties related to non-linear bias and intrinsic alignments \citep{Pantos_2026}. By considering these small-scale systematics, future studies will be able to extend such model-independent measurements to the scale-dependent Weyl potential $\hat{J}(k, z)$, offering more detailed tests of modified gravity.

\normalem
\begin{acknowledgements}
This work is supported by the National Key R\&D Intergovernmental Cooperation Program of China (2024YFA1611500,2023YFE0102300), the National Natural Science Foundation of China (12173078).
The analysis code and chains are available at \url{https://github.com/irosphis/cosmosis_weyl} and archived on Zenodo at \url{https://doi.org/10.5281/zenodo.22220484}.
\end{acknowledgements}

\bibliographystyle{raa}
\bibliography{bibtex}

\end{document}